\documentstyle[psfig]{mn}
\title[Morphology of 9 Radio-selected Faint Galaxies]
{The Morphology of 9 Radio-selected Faint
Galaxies from deep HST Imaging} 
\author[N.D. Roche, J.D. Lowenthal, D.C. Koo]{Nathan D.
Roche$^{1,4}$, James D. Lowenthal$^{2,5}$ and David C. Koo$^{3,6}$\\
$^1$Institute for Astronomy,
     University of Edinburgh,
     Royal Observatory, 
     Edinburgh EH9 3HJ,
     Scotland.\\
$^2$Department of Physics and Astronomy,
     University of Massachusetts,
     Amherst, 
     MA 01003-4525,
     USA.\\
$^3$UCO/Lick Observatory,
       Department of Astronomy and Astrophysics,
      University of California,
     Santa Cruz,
       CA 95064, 
       USA.\\
{$^4$ \verb"ndr@roe.ac.uk"}\hspace{8mm}   
{$^5$ \verb"james@velo.astro.umass.edu"}\hspace{8mm}
{$^6$ \verb"koo@ucolick.org"}\hspace{8mm}
}

\begin{document}

\maketitle
\begin{abstract}
Using  the   HST  WFPC2  we   perform  deep  $I$-band  imaging   of  9
radio-selected ($\rm F(8.5~GHz)\geq 14\mu \rm Jy$) faint galaxies from
the  Roche, Lowenthal and  Koo (2002)  sample. Two are also
observed in $V$ using HST STIS. 

Six of the galaxies have  known
 redshifts, in the range $0.4<z<1.0$.
 Radial intensity profiles indicate that 7 are disk
 galaxies and 2 are bulge-dominated. Four of the six with redshifts have a
 high optical  surface brightness in comparison with previous studies
 of disk galaxies at similar redshifts (e.g. Lilly et al. 1998).  
The HST imaging reveals that 2 
of the 9 galaxies are in close interacting pairs and
 another 5 show morphological evidence of recent interactions -- 
two are very asymmetric ($A_{asym}\sim 0.4$) and,
 three have large, luminous rings resembling the collisional starburst
 rings in the Cartwheel
 galaxy. For the two ring galaxies
 with redshifts, we measure ring radii of 7.05 and 10.0
 $h^{-1}_{50}$ kpc, which suggest post-collision ages 0.1--0.2 Gyr. 
One has a fainter inner ring, like the original Cartwheel. 
The remaining two appear to be late-type  barred galaxies  and 
 relatively undisturbed.

Our HST imaging confirms the high
 incidence of interactions and dynamical disturbance in 
 faint  radio-selected galaxies, as  reported by e.g.   Windhorst et
al.  (1995) and  Serjeant. et  al.  (2000).  In the great majority of  these
galaxies the high radio luminosities are probably  the result of
interaction-triggered starbursts. However, one interacting 
galaxy is a very radioluminous
 giant elliptical, with red $V-I$
colours, a normal surface  brightness  and no  evidence of  star-forming
regions, so its radio source is probably an obscured AGN. 
The mixture of observed morphologies suggests
that enhanced radio luminosities often persist to a
 late stage of interaction,
i.e. at least   $\sim 0.2$  Gyr  after the  perigalactic
encounter.
\end{abstract}


\section{Introduction}
The radio emission from star-forming galaxies is of great
importance as an  apparently
unbiased tracer of the star-formation rate (SFR), being neither
obscured nor enhanced by the presence of dust (e.g. Condon 1992).
 At sub-milliJansky
fluxes, deep radio surveys detect large number of star-forming
galaxies, which overtake radio galaxies and other active galactic
nuclei (AGN) as the most numerous type of source. Hence a  deep
radio-selected source sample will efficiently select
the most intensely star-forming galaxies in the survey volume.

Hammer et al. (1995) investigated  the optical IDs of a sample of 
radio sources
 detected in  a deep  Very Large
Array  (VLA) survey of one of the  Canada-France
Redshift  Survey fields, to a limit in integrated flux 
 $F(5.0~\rm
GHz)\geq  16\mu Jy$. Spectroscopy revealed these galaxies to be  mixture
of  (i) giant ellipticals,  presumably  with AGN,  (ii)  star-forming 
disks at redshifts out to $z=1.16$, and
(iii) lower redshift, high-excitation emission-line galaxies.
Windhorst et al. (1985) and Richards et al. (1998) 
obtained similar findings -- although with fewer AGN and more disk
galaxies -- at even fainter flux limits. Using HST imaging, they
 estimated that at least  $\sim 60$
 per cent  of the radio-selected galaxies 
were interacting.

In a previous paper (Roche, Lowenthal and Koo 2002; hereafter
RLK02) we identified the 
most probable optical IDs for 50 out of 51 $F(8.5~\rm GHz)=7$--900
$\rm \mu Jy$ sources, detected in
three deep VLA surveys.
These galaxies were imaged  in $BRI$ using the Keck Low
Resolution Imaging Spectrograph (LRIS), and in $K$ with the Near
Infra-red 
Camera (NIRC).
Spectra and redshifts were obtained 
for 17, using LRIS in spectroscopic mode. Of these, 14 were found to be emission-line
galaxies at redshifts up to $z\sim 1$, and the other 3 were QSOs at $z>1$.
 A further 9 of the radio ID galaxies had known redshifts from previous
observations, giving a total of 26 with known
redshifts. 
RLK02 concluded that more than half of 
these 26  were disk
galaxies with enhanced radio luminosities, resulting from major starbursts
($\rm SFR\sim 100 M_{\odot}yr^{-1}$). The remainder were 
apparently normal, non-starburst 
 spirals and ellipticals at
lower redshifts ($z<0.4$) ($\sim 20$ per cent), QSOs ($\sim 15$ per
cent), or giant radioluminous ellipticals suspected to contain 
obscured AGN ($\sim 8$ per cent).

The LRIS spectra of the 14 non-QSO  galaxies 
showed  the expected emission lines, 
 $\rm [OII]3727\AA$, $H\beta$ and $\rm [OIII]5007\AA$, but the
 line luminosities typically correspond to
 SFRs an order of magnitude lower than the radio luminosities imply.
Other surveys of radio-selected galaxies 
have found a similar discrepancy between 
emission-line and radio fluxes  (e.g. Smith et al. 1996;  Beck, Turner
 and Kovo  2000; Serjeant et al. 2000). The simplest explanation
is a high dust extinction ($A_V\simeq 2$--3 mag) of the starburst
regions.
We also found that  11 of the 26 galaxies  were in  close  pairs and several
 others looked
 disturbed. However, it was clear that to adequately 
investigate the interaction status of these galaxies,  much higher
resolution imaging, e.g WFPC2,
 would be required. 

Serjeant et al. (2000) performed
 an WFPC2 imaging 
survey of radio-selected star-forming galaxies, and describe the first four. 
These galaxies are at lower redshifts, $z\sim 0.2$, than the RLK02
 sample but have  
comparable radio luminosites. They found 
at least one galaxy to be  interacting and all 4 to be
 disturbed to some degree, and 
quantified 
 morphological disturbance in terms of a
rotational asymmetry parameter $A_{asym}$ 
(Conselice, Bershady and Jangren 2000). They concluded that
radio-selected
galaxies are generally more asymmetric than  
             optically-selected galaxies at similar magnitudes.
 Similarly, and at higher redshifts, 
Fomalont et al. (2002) identified 37/63 faint 
 $\rm F(8.5~GHz)\geq7.5\mu \rm Jy$ radio sources with $I\leq 23.3$ 
galaxies, and
using WFPC2 imaging found a high proportion, 46 per cent, to be
multiple or interacting.                                                            

In this paper we perform a similar WFPC2
              study
 of a subsample of the RLK02 radio IDs.
Our HST program was allocated 
a total of 24 orbits, which was devoted to WFPC2
 $I$-band imaging of four fields, containing
 a total of 9 of the RLK02 galaxies. For the first two fields we
imaged in  parallel with STIS, and obtained images in the broad STIS
              $V$-band of two of the galaxies.
Section 2 of this paper describes the observations and data
reduction. Section 3 catalogs the observed galaxies and 
presents HST images.
In Section 4, radial intensity
profiles are presented, model
profiles fitted, and   surface brightness compared 
 with local galaxies.
In Section 5 the evidence for interactions is investigated.
Asymmetry parameters are evaluated and 
model profiles subtracted from the galaxies to highlight `residual'
features. In Section 6, galaxy colours are interpreted.  Section 7
is a discussion of the nature of these galaxies and the source of
              their radio luminosity.
\section{Data}
\subsection{Sample Selection}
The galaxies studied here are an unbiased subsample of the
galaxies identified by RLK02 on Keck LRIS images as probable
optical counterparts of the radio detections on two VLA surveys.
The first survey is of the SA68 field, centred at R.A. $23^h 59^m
15^s$  Dec 14:55:00 (equinox 2000.0 used throughout),
 in which observations  at 5.0 GHz detected
 sources to integrated flux limits of  $60\rm \mu Jy$, or
$\sim 40$--$45\rm \mu Jy$ at 8.44 GHz (Weistrop et
al. (1987). The WFPC2 field of view covers only
one of these sources at a time. 

The second survey, smaller in area but much deeper, is
of the  `Lynx2' or `16V' field ( R.A. $08^h 45^m 04^s$ 
Dec 44:34:05), observed at 8.44 GHz for 63
hours in December 1989 and January 1990. This gave a noise level of
$\sigma=3.21\rm \mu Jy/beam $ and a point-source completeness
limit at $F(8.44~{\rm GHz})\simeq 14.5\rm \mu Jy$ -- Windhorst et
al. (1993) catalog the detections. A
 WFPC2 field may include  several of these fainter sources.

\subsection{HST Observations}
The HST observations, in 14 time slots from  September 2000
to September 2001, consisted of two pointings within the SA68 VLA
survey (Weistrop et al. 1987)
and two in the Lynx2/16V (Windhorst et. al 1993) field.
 Our program originally called for observations  
with both WFPC2 and
STIS, but the temporary failure of STIS partway
through our observations meant that we used only WFPC2 for the targets
in the Lynx2 field.
Imaging was performed in four telescope pointings as
	described below.

(i) The optical ID for the radio source SA68:10 (hereafter
	`S10') was centered in the WFPC2 and imaged in the $I$-band
	(F814W) for 5 orbits, giving 10 exposures of 1200s or 1300s
	for a total of 12.4 ksec.  Each orbit was split into two
	exposures for cosmic-ray rejection, and the telescope was
	dithered a few arcseconds between successive orbits to
	improve flat-fielding and bad pixel rejection.  STIS,
	operating in parallel, simultaneously imaged another galaxy
	from the same survey, SA68:4 (hereafter `S4'), for 5 exposures
	of 400 sec each totalling 2000 sec in a wide $V$-band filter
	('mirvis').  For the remainder of the available time at that
	pointing, STIS was used in spectroscopic mode, but  the
	resulting spectra were of poor signal-to-noise and
        will not be discussed
	here further.

(ii) The telescope was rotated 180 degrees at the same
	pointing as in (i) to switch instruments, so that the WFPC2
	was centered on S4 and STIS on S10.  Again we obtained 10
	exposures totalling 12.4 ksec in F814W with WFPC2 and 5 imaging
	exposures totalling 2.0 ksec in STIS.

	(iii) The Lynx2 survey source 16V31 was centered in the WFPC2
	and observed in the $I$-band for 7 orbits, giving 14 exposures
	totalling 17.4 ksec.

	(iv) As (iii) but centered on 16V22.

The 
observed area of the 16V31 pointing also included the sources
16V25, 16V30 and 16V34. The 16V22 pointing overlapped slightly with
 the third and again included 
16V25, together with 16V21 and 16V26. Hence in total we observe 9
radio-selected galaxies.
  
\subsection{Data Reduction}
The data were downloaded from the STScI archive already flat-fielded
and calibrated, and further reduction was performed using {\sevensize
IRAF}, as described in  more detail by 
 Roche, Lowenthal and Woodgate (2000). 
The WFPC2 has 4 chips (c1--c4), each $800\times 800$
pixels,
with a pixelsize 0.046 arcsec on c1 (the PC) and 0.0966 arcsec on
c2--c4.
 The STIS CCD produces a single $1024\times 1024$ pixel
image, pixelsize 0.05077 arcsec. 

Cosmic rays were
removed from WFPC2 data using either the `nukecr' routine developed by 
Luc Simard or `bclean' on Starlink Figaro, and cosmic ray pixels were
flagged on the data quality file. The STIS data 
were supplied with cosmic rays already rejected and flagged. 

The exposures had been spatially dithered with offsets of up to a few arcsec
between each orbit. Offsets were measured  using the {\sevensize
IRAF} routines `precor', `crossdriz' and `shiftfind`. The WFPC2 and STIS 
exposures (together with their data quality files) were rebinned to a
finer $2048\times 2048$ grid, using {\sevensize
IRAF} `drizzle', and positionally registered using the
measured offsets (rebinning scale factors were 0.42 for
WFPC2 chips 2--4, 0.45 for c1 and 0.53 for STIS). 
The drizzled exposures were then stacked using IRAF `imcombine',
 with rejection of those pixels discrepant by $>3\sigma$ and
those flagged as `bad pixels' or as cosmic rays.
Finally the stacked exposures were trimmed to remove the low-quality 
edges.
\subsection{Detection and Photometry}
Sources were detected on the drizzled and stacked images using
SExtractor (Bertin and Arnauts 1996), with a detection threshold of
$2\sigma_{sky}$ in a minimum area of 16 rebinned pixels
for WFPC2 c2--c4, 36 for c1, and  25 for STIS.
Over 200 galaxies are detected on each large WFPC2
chip.
SExtractor provides `total' (Kron-type) magnitudes,
derived by fitting a series of elliptical
apertures to each detection. We have found that these
 `total' magnitudes are usually
reliable for WFPC2 data (they agree closely with the magnitudes from
{\sevensize STIS} `isophote.ellipse' at the largest radii), and 
hence we adopt these here as the true $I_{814}$ 
magnitudes.

Magnitudes are given throughout in the AB system, in which,
 in any passband,  $m_{AB}=-2.5~{\rm log}_{10}~F_{\nu}-48.60$,
 where 
$F_{\nu}$ is flux in ergs $\rm s^{-1} cm^{-2} Hz^{-1}$.
The stacked images are normalized to the mean exposure time 
(1240 sec for the WFPC2 SA68 fields, 1243 sec for the
WFPC2 Lynx2 fields, and 400 sec for STIS), and 
photometric zero-points  derived as
$$z.p.(AB)=-2.5{\rm log}~({photflam\over
exptime}).{\lambda_{pivot}^2\over c}-48.60-A_{gal}$$
where
`photflam' is as given in the exposure headers,
$\lambda_{pivot}=7995.9\rm \AA$ for F814W and $5835.5\rm \AA$ for
STIS $V$), and  $A_{gal}$ is the Galactic extinction (
from the NASA/IPAC Extragalactic Database).



 The $
2\sigma$ thresholds are approximately 24.7 (SA68) or 25.0(Lynx2)  $I$ mag $\rm arcsec^{-2}$
for WFPC2 and  24.2 $V$ mag $\rm arcsec^{-2}$ for STIS imaging.
All  9  of the radio-selected galaxies in the observed region are detected.

\section{The Observed Sample: Catalog and HST Images}
\begin{table}
\caption{Radio-selected galaxies observed with WFPC2/STIS; positions (J2000),
HST $I$ magnitudes (errors $\leq \pm 0.04$ mag),
 ground-based $B-I$ colours from RLK02 (errors $\sim\pm
0.1$--0.3), redshifts (RLK02), integrated 8.44 GHz fluxes, estimated
luminosities in the restframe 8.44 GHz
(RLK02) and the corresponding SFRs, with the lower and upper values
corresponding to  the $L_{rad}$-SFR relations of 
Carilli (2001) and Condon (1992). These SFRs also
assume that all radio emission is due to star formation and do not apply if
there is a large AGN contribution.}
\begin{tabular}{lcccc}
\hline
Radio & \multispan{2} Optical position & $I_{814}$  & $B-I$ \\
\smallskip
source      & R.A.  & Dec.&    &   \\
S4 &   00:17:41.75 & 15:50:03.2 & 20.84  & 2.46  \\
S10 & 00:18:03.66 & 15:49:04.6 & 22.22  & 0.85  \\
16V21 & 08:44:58.22 &  44:33:13.46 & 23.05 & 0.50   \\
16V22 & 08:44:59.00 &  44:33:44.00 & 22.36 & 0.59    \\
16V25 & 08:45:05.48 & 44:34:15.14 & 21.71 & 3.02   \\
16V26 & 08:45:05.61 &  44:33:56.56 & 24.49 & 0.16   \\
16V30 & 08:45:07.91 & 44:33:51.60 & 20.28  & 1.96   \\
16V31 & 08:45:08.35 & 44:34:38.58 & 20.74 & 1.96 \\
16V34 & 08:45:14.33 & 44:34:50.97 & 19.78 & 1.38  \\
\hline
\end{tabular}
\begin{tabular}{lcccc}
\hline
Radio & Redshift & 8.44 GHz flux &  ${\rm log}~L_{8.44}$ &
 ${\rm SFR}(L_{8.44})$  \\
\smallskip
source &  &  ($\rm \mu Jy$) & ergs $\rm s^{-1}$ & $M_{\odot}\rm yr^{-1}$ \\
S4 & 0.7125  & $270.\pm 20.^{[1]}$  & 40.91 & 953--2239 \\
S10 & 0.9923  &  $53.\pm 12.^{[1]}$  & 40.54 & 406--955 \\
16V21 & ? &  $50.4\pm 5.5$ &  (38.96--$40.51)^{[2]}$
 & (10.6--$884)^{[3]}$  \\
16V22 & 0.4259  &  $20.6\pm 3.6$ &  39.29  & 22.9--53.7 \\
16V25 & 0.7273 &  $31.5\pm 3.6$ & 39.99 & 114--269 \\
16V26 & ? &   $35.3\pm 3.6$ & (38.80--$40.35)^{[2]}$  & (7.3--$612)^{[3]}$ \\
16V30 & ?  & $13.8\pm 3.4$ & (38.39--$39.94)^{[2]}$ & (2.9--$238)^{[3]}$\\
16V31 & 0.4977  & $62.4\pm 5.1$ & 39.92  & 97-229 \\
16V34  & 0.4257 & $31.0\pm 5.9$ & 39.47  & 34.6--81.3  \\
\hline
\end{tabular}
$^{[1]}$ estimated from the 5.0 GHz flux (Weistrop et al. 1987)
using the Condon (1992) SED. For galaxies without known
redshifts, ${[2]}$ is
the range of radio luminosities estimated for the redshift
range 0.2--1.0; ${[3]}$ is the range of SFR estimates from that for
$z=0.2$ and the Carilli (2001) relation to that for $z=1$ and the
Condon (1992) relation.
\end{table}

Our HST sample therefore consists of 9 radio-selected galaxies, of which
the Keck LRIS spectroscopy of 
RLK02 provides reliable redshifts for 6. 
Table 1 gives the optical positions of the 
 galaxies, the redshifts where known, the WFPC2 $I$ magnitudes and
radio fluxes, and  
the RLK02 estimates of the restframe  8.44 GHz luminosities (in the form 
$\nu L_{\nu}$) and corresponding SFRs.
 All luminosities and SFRs throughout are given
for $H_0=50$ km $\rm s^{-1} Mpc^{-1}$ and a flat $\Omega_{M}=0.3$, 
$\Omega_{\Lambda}=0.7$ cosmology. 

The SFRs are estimated  assuming no AGN contribution to the radio
flux (see Section 7.1).  The lower and upper values correspond respectively to  the
$L_{rad}$-SFR relations of Carilli (2001) and Condon (1992). These
relations give the SFR for $M>5M_{\odot}$ strs, but here we give SRFs as
totals for all masses of star, assuming  
an Initial Mass Function (see RLK02) of  $x=2.35$ (where ${d{\rm
N}\over d  {\rm M}}\propto \rm M^{-x}$) at  $\rm 0.7<M<120 M_{\odot}$,
and $x=1.3$  at $\rm 0.1<M<0.7  M_{\odot}$. The total SFRs  would be a
factor 1.66 higher for the Salpeter IMF.

 RLK02 give details and plots of the spectra;  important properties are briefly
 described below.
`W' refers to a line equivalent width in the restframe (the observed value 
divided by $1+z$). We have no redshifts or spectra for 16V21, 16V26 or 16V30.

 16V22 has strong emission, $\rm W([OIII]5007)=30.2\rm
\AA$, and a high excitation ($F({\rm [OIII]})/F({\rm H\beta})$) ratio of 
$2.93\pm 0.14$,  consistent with  an HII galaxy. 16V25 has 
moderate emission lines typical of a normal spiral, with 
$\rm W([OII]3727)=8.8\AA$, as do 16V31 and 16V34, with $\rm
 W([OIII]5007)=6.3\AA$
and  $3.4\AA$ respectively.  For 16V31 and 16V34 we have
 excitation
 ratios, $0.68\pm 0.09$ and $0.54\pm
 0.06$ respectively, consistent with late-type or starbursting spirals but
 not AGN (unless obscured).   16V31 also has strong
$\rm H\delta$ absorption of $\rm W=9.1\AA$, classifying it as an `e(a)'
galaxy (Poggianti and Wu 2000).
S4, classed by RLK02 as an elliptical,
has moderate emission, $\rm W([OII]3727)=7.5\AA$, but
 S10 has 
$\rm W([OII]3727)=52.5\AA$, indicating a
major starburst.

The spectral line profiles also provided estimates of  internal and rotation
 velocities, given in RLK02. 16V22, the only
sub-$L^*$ galaxy in this sample, had low velocities
consistent with a HII galaxy. The other five spectra all 
show evidence of rotation ($v_{rot}\sim 200$ km $\rm s^{-1}$),
 consistent with $\geq L^*$ spirals.

 Figure 1 shows greyscale
plots of the galaxies as observed with  WFPC2 and STIS. The HST imaging
reveals
 many significant details not apparent in the 
ground-based observations (RLK02 Figure 1). S4 has a fainter companion,
S10 and 16V21 are both highly asymmetric with multiple bright knots
(16V21 may be a merging system). 16V22 and 16V25 appear to be
relatively undisturbed barred galaxies, with comparison by  eye
with local galaxy catalogs (e.g. Sandage and Bedke 1998) suggesting
 approximate types SBm and SBbc
respectively.  16V26 is faint but obviously disturbed, and 16V30,
16V31 and 16V34 all have large, bright rings. These features are discussed
further in
 Section 5.

\section{Intensity Profiles}
\subsection{Disk and Bulge Fitting}
We investigated the radial intensity profiles of these galaxies
using 
{\sevensize IRAF} `isophotes', which, with
 centroids, ellipticities and position angles
from the SExtractor
catalog used as starting parameters, 
fits each
 galaxy with a series of
elliptical isophotes, providing a radial intensity profile on the
semi-major axis. The resulting profiles were then
best-fitted with exponential, $I_0~{\rm exp}(-r/r_{exp})$, and
bulge, $I_0~{\rm exp}(-7.67[r/r_{e}]^{0.25})$,
 functions.

 Figure 2 shows the $I$-band profiles of the 9
galaxies, together with the best-fitting (lowest rms residuals) 
models and their $r_e$ or $r_{exp}$ 
scalelengths. Table 2 gives the scalelengths and central surface
 brightness (SB) of
the best-fit models together with the observed central SB, averaged 
over the pixels at $r\leq 0.1$ arcsec.

S4 is closely fitted by a bulge model, with the excess at
$1.5<r<2.5$ arcsec being due to a companion galaxy. The ring around
16V30 
gives the impression of a spiral, but a bulge model is adopted here as 
it is  a much better fit at $r<0.8$
arcsec. 
 The other 7 galaxies are much closer to purely 
exponential disks, but some have large ($\sim 0.2$
mag) residuals.

\begin{table}
\caption{Best-fitting radial intensity profiles:
profile type, $r_{exp}$ or $r_e$ in arcsec,
and central ($r=0$) SB of the fitted model profile, together 
with  the observed  central ($r<0.1$ arcsec) SB.}
\begin{tabular}{lcccc}
\hline
Galaxy & Type &  $r_{exp}$ or $r_{e}$ & \multispan{2} Central SB
($I$ mag $\rm arcsec^{-2}$)\\
   &  & arcsec  & Model & Observed  \\ 
S4 & bulge & $1.46\pm 0.09$ &  $15.70\pm 0.10$ &
$20.21\pm 0.03$ \\  
S10 & disk & $0.61\pm 0.02$ & $22.10\pm 0.03$ & 
$22.30\pm 0.08$ \\
16V21 & disk & $0.52\pm 0.02$ &  $23.07\pm 0.05$ & $23.20\pm
0.12$ \\
16V22 & disk & $0.54\pm0.01$  & $22.00\pm 0.03$ & 
$22.04\pm 0.12$ \\ 
16V25 & disk & $0.42\pm 0.02$   & $21.38\pm0.06$
& $21.63\pm 0.06$  \\
16V26 & disk & $0.27\pm 0.01$ &  $23.10\pm 0.07$ & $23.13\pm
0.12$ \\
16V30 & bulge & $2.85\pm 0.12$  & $16.11\pm0.05$ & $19.04\pm
0.04$  \\
16V31 & disk & $0.72\pm 0.02$  & $20.80\pm 0.04$
& $20.86\pm 0.04$ \\
16V34 & disk & $1.26\pm 0.04$  & $20.60\pm
0.04$ & $20.72\pm 0.04$ \\  
\hline
\end{tabular}
\end{table}

\subsection{Surface Brightness}
 RLK02 
estimated rest-frame blue absolute magnitudes, $M_B$, and 8.44 GHz 
luminosities, $L_{8.44}$, for the radio IDs with known redshifts.   
Firstly, a grid of model spectral energy distributions,
representing galaxy types from ellipticals to pure starbursts over a
wide range of redshifts, 
was  generated using `Pegase2' (Fioc and Rocca-Volmerange 2000). 
For each model SED, a redshift-dependent 
correction $(B_{rest}-R_{obs})(z)$ 
was calculated. The observed colour of each galaxy ($B-R$, $B-I$
and/or $I-K$) was compared with
the models at the  same redshift, and interpolation between
the two models closest in colour then gave an estimated 
 $(B_{rest}-R_{obs})(z)$, which was used to derive  $M_B$ from
the apparent $R$  magnitude.  $L_{8.44}$ was similarly estimated from the
observed 8.44 GHz fluz assuming the Condon (1992) model radio SED. 

Using the more accurate WFPC2 photometry, we re-estimated $M_B$
for the 6 galaxies with known redshifts. For each galaxy, the
correction 
$(B_{rest}-I_{814})(z)$ is derived from  the same interpolated model
SED as
used
by
 RLK02, and then used to derive  $M_B$ from the
WFPC2 $I$ magnitude (Table 3).

The surface brightnesses and luminosities of
        our radio-selected galaxies are compared with
 those of normal
        optically-selected field galaxies. 
 The half-light radius $r_{hl}$
is equal to $r_{e}$ for bulges and $1.679~r_{exp}$ for disks.
Binggeli, Sandage and Tarenghi (1984) find that for nearby E/S0s with 
$M_B\leq -20$ (for $H_0=50$ km $\rm s^{-1}Mpc^{-1}$)
$${\rm log}(r_{hl}/{\rm kpc})=-0.3(M_B +18.62)\eqno{1}$$
and for the less luminous ($M_B> -20$) E/S0s
$${\rm log}(r_{hl}/{\rm kpc})=-0.1(M_B +15.75)\eqno{2}$$
For spirals, the Freeman (1970) central SB is $\mu_B=21.65$ mag 
$\rm arcsec^{-2}$ at all luminosities, which  corresponds to 
$${\rm log}(r_{hl}/{\rm kpc})=-0.2M_B -3.16.$$

However, more recently, Cross and Driver (2002) determined
 a bivarate brightness function for
 45000 disk galaxies in the 2dFGRS, and found evidence of a
positive correlation
 between SB and luminosity. Their best-fit relation
 corresponds to
$${\rm log}~(r_{hl}/{\rm kpc})=-0.144M_B -2.034\eqno{3}$$ 
with scatter $\sigma({\rm log}~r_{hl})=0.103$, or to a central SB
$$\mu_B=21.326+0.28(M_B+21.23)\eqno{4}$$
\onecolumn
\begin{figure}
\includegraphics{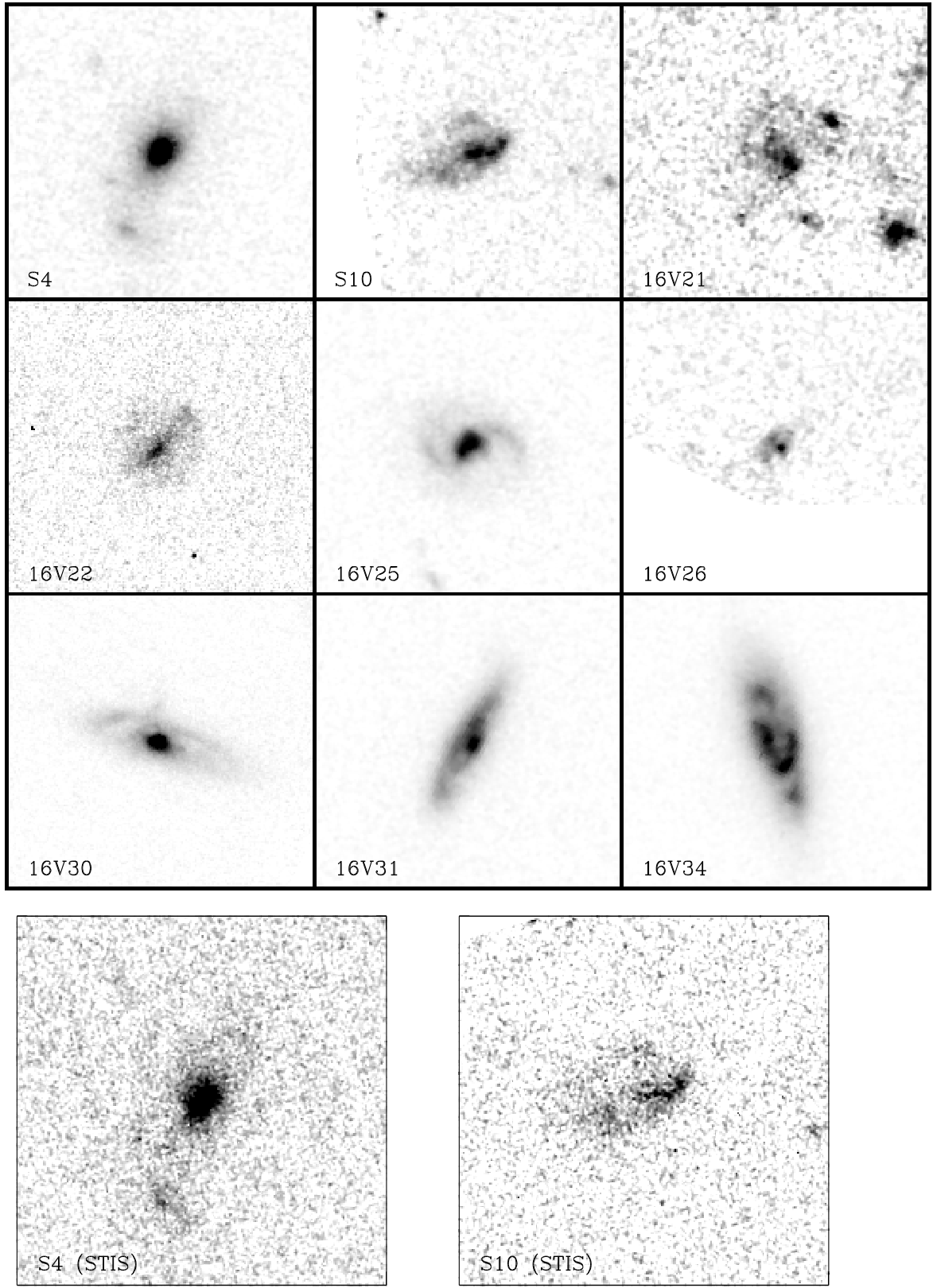}
\vskip22cm
\caption{Greyscale plots of $6.7\times 6.7$ $\rm arcsec^2$ areas 
centred on each of the sample galaxies, as observed in $I_{814}$ with
WFPC2 (above), and the $V$ band with STIS (bottom). Oriented with
North at the  top, East at the  left. Greyscales are
linear with white the sky background, and the black levels set by eye
for each galaxy depending
on  its surface brightness.}      
\end{figure}
\twocolumn
\onecolumn
\begin{figure}
\psfig{file=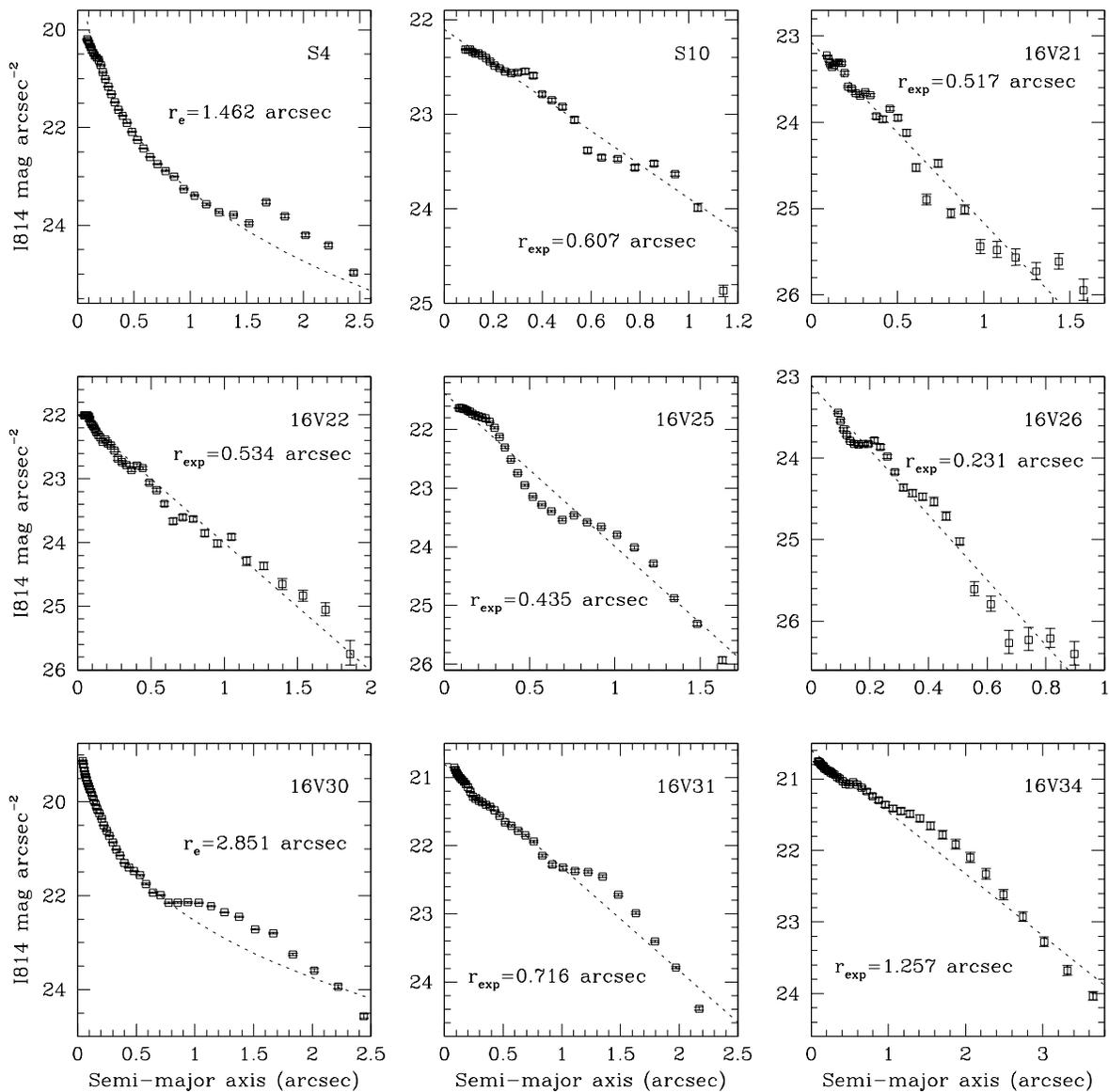,width=170mm}
\caption{Radial profiles of the 9 radio detected galaxies, from WFPC2
$I$-band
observations. Dotted lines show the best-fit disk or bulge profiles,
with $r_{e}$ or $r_{exp}$ scalelengths as indicated.}
\end{figure}
\twocolumn
with scatter $\sigma(\mu_B)=0.517$ mag.

On Figure 3, our 
$r_{hl}$ and luminosity ($M_B$) estimates for  the
 radio-selected galaxies are plotted 
with the $r_{hl}-M_B$ relations from Equations 1, 2 and 3
For each galaxy we calculated 
$\Delta(M_B)$ as the difference between the observed  $M_B$ and that
given by   the appropriate relation
 (i.e. Equation 1 for S4 and
 16V30, Equation 3 for the other four) for the observed $r_{hl}$.
This provides a direct comparison between the
rest-frame blue-band SB of our radio-selected galaxies and that of
normal field galaxies.

However, this 
is likely to be  be an underestimate 
for spirals seen close to edge-on (i.e. 16V31
and 16V34), which would require corrections for
internal extinction. Hoerver, the  central
SB (Table 2), should be much less sensitive to the inclination
angle, to local disk galaxies.
 The $B$-band 
 central SB, corrected to $z=0$, can be estimated as
$$\mu_{B0}=\mu_{I(z)}+(B_{rest}-I_{814})_{z}-10~{\rm log}~(1+z)$$
For the disk galaxies, we calculated  $\Delta\mu_B$ as the 
 the difference between the observed $\mu_{B0}$ and the central SB given by
Equation 4 for the $M_B$ given by Equation 3 for the observed $r_hl$
($not$ for the observed $M_B$).
 We see that, for the face-on 16V25, $\Delta M_B$ and
$\Delta\mu_B$ are very similar, but $\Delta\mu_B$ is much
greater for the edge-on 16V31 and 16V34. 

\begin{table}
\caption{For the 6 galaxies with known redshifts -- 
absolute magnitude in the rest-frame $B$-band 
($+2.5~{\rm log}~h_{50}$); the best-fitting scalelength in  $h_{50}
^{-1}$ kpc;
$\mu_{B0}$, the  central  SB
corrected to rest-frame $B$ at $z=0$; and the two  estimators 
described in the text of the rest-frame blue SB
 relative to average  local galaxies of the same type, $\Delta M_B$ and
$\Delta\mu_B$, in  rest-frame $B$
 magnitudes. Likely errors $\sim 0.1$ mag for $M_B$ and  $\mu_{B0}$,
$\sim 0.14$ mag for $\Delta M_B$, $\sim 0.2$ mag for $\Delta\mu_B$.}   
\begin{tabular}{lccccc}
\hline
\smallskip
Galaxy & $M_B$ & $r_{exp}$ or $r_{e}$  & $\mu_{B0}$  &  $\Delta(M_B)$ & $\Delta\mu_B$  \\
S4    & -22.78 & $14.75\pm 0.91$  & 18.19 & -0.26 & - \\
S10  & -22.68  & $6.804\pm 0.247$ & 19.22 & -1.21 & -2.04 \\ 
16V22 & -19.89 & $4.232\pm 0.063$  & 20.83 & +0.15 & -0.83 \\       
16V25 & -21.98 & $4.230\pm 0.153$ & 19.53 & -1.94 & -2.13 \\
16V31 & -21.57 &   $6.124\pm 0.166$ & 19.78 & -0.42 & -1.57 \\
16V34 & -22.32  &  $9.844\pm 0.274$ & 19.66 & +0.26 & -1.29 \\
\hline
\end{tabular}
\end{table}
 On the basis of these studies, the central SB of 16V22 is 
near the average for a disk galaxy of its size and redshift, whereas   
16V25, 16V31, 16V34 and S10 are $\sim 1$ mag above the average. All
four lie at about the top of   the observed 
SB range of  CFRS/LDSS galaxies in the $\mu_B-z$ plot of Lilly et
al. (1998).

\begin{figure}
\psfig{file=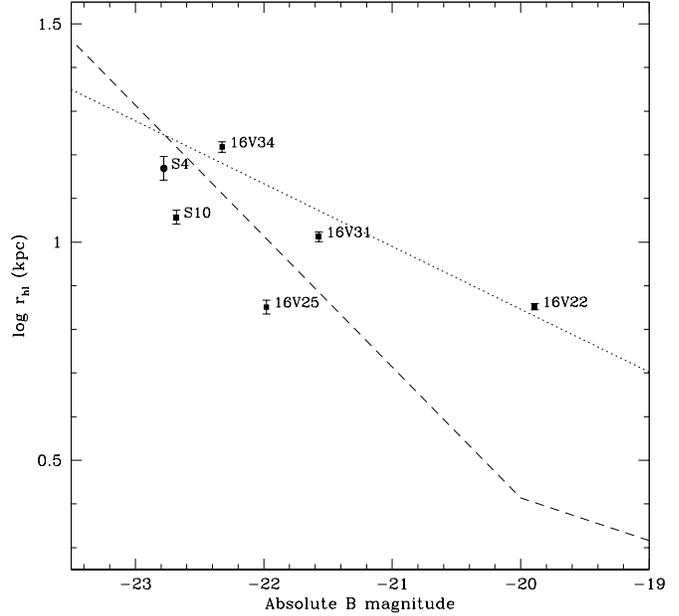,width=90mm}
\caption{Half-light radius, $r_{hl}$, against restframe $B$-band
absolute magnitude, for $H_0=50 $ km $\rm s^{-1}Mpc^{-1}$, of the 6
radio-selected galaxies with known redshifts, compared with
size-luminosity relations for $z\sim 0$ spirals (Equation 3; dotted) and 
 ellipticals (Equation 1 and 2: dashed).}
\end{figure}
 The relatively high SBs of these galaxies may then  favour  the interpretation
of their enhanced radio luminosity as the result of starbursts. 
 In all four, the high SB appears to be associated with the whole
disk or with extended features, such as the bright rings of 16V31 and
16V34 (see below), and not with any central point-source.
 16V22 has a more moderate SB, but its blue colours would be evidence
for starbursting. However there is no indication that S4 is
starbursting as its SB and red colours (Section 6) are consistent with a
passively evolving elliptical. 

\section{The Evidence for Interactions}
\subsection{Interacting Pairs}
Two of the 9 galaxies appeared to be interacting with smaller
companions, although we have no definite (i.e. with spectroscopic
kinematics) confirmation of this. S4 appeared to be an isolated galaxy
in the RLK02 data, but the HST reveals a much fainter
(the demerged SExtractor magnitude is $I=22.67\pm 0.02$),
disk-type companion 1.92 arcsec to its SSE.
The luminosity ratio is $0.2:1$ and projected separation
$19.4 h_{50}^{-1}$ kpc. S4 is not greatly disturbed but 
 does show  isophotal twist in the central $r<0.4$ arcsec.
  16V21 has bright knots suggesting active star formation, and is very
  asymmetric. It appears to be  connected by a luminous filament to a
 smaller, high-SB companion 3.24 arcsec to the
SW, with $I=24.05$ (luminosity ratio 0.4:1).
The other 7 galaxies have no obvious companions,  but we investigate
below the possibility that they show evidence of very 
recent interactions.
\subsection{Asymmetry Parameters}
Conselice, Bershady and Jangren (2000) presented a quantitative measure
of rotational asymmetry, which can be used to distinguish between
`normal', undisturbed
galaxies and those undergoing an interaction or merger. The statistic
they found to be most successful is evaluated as
$$A_{asym}=\frac{\Sigma|I_{0}-I_{180}|}{2\Sigma|I_{0}|}-k$$
where $\Sigma$ represents a summation over the galaxy image, $I_{0}$
the image of the unrotated galaxy, $I_{180}$ the image of the same 
galaxy rotated by
$180{\degr}$ about a central point (which is fine-tuned  by 
iteration to  minimize the resulting  $A_{asym}$), and $k$ an offset produced by background
noise, which is corrected for 
 by evaluating $A_{asym}$ on a blank region of the field. A perfectly
symmetric galaxy would give $A_{asym}=0$ while $A_{asym}=1$ is the 
maximum possible. Conselice et al. (2000) found this 
 statistic to be most reliable when
evaluated within a radius where a `Petrosian function', $\eta(r)$, falls
to 0.2. This function is defined as
$\eta(r)={I(r)\over \langle I(<r)\rangle}$ where  $I(r)$ is intensity at
radius $r$ and $\langle I(<r)\rangle$ the mean intensity within radius
$r$ (Petrosian 1976; Bershady, Lowenthal and Koo 1998). 
Typically, $\eta(r)=0.2$ at $r\sim3r_{exp}\sim$ 
the edge of the visibly luminous galaxy. 

 Undisturbed ellipticals and early-type spirals are generally found to
have
$A_{asym}<0.1$, late-type spirals and irregulars $A_{asym}\sim
0.1$--0.2, while galaxies undergoing major mergers 
 have higher asymmetries of
 $A_{asym}\sim 0.25$--0.6. 
Serjeant et al. (2000) calculated $A_{asym}$ for their four
$z\sim 0.2$ radio
selected galaxies as 0.47, 0.112, 0.185 and
0.317. The mean is 0.27 and the dispersion 0.16.

Using an {\sevensize IRAF} routine written  by
 M. Bershady and C. Conselice, we evaluated 
 $A_{asym}$ (and associated error) for the 9  galaxies in our sample.
(Table 4). These are at somewhat higher redshufts.
Conselice et al. (2000) modelled the effect of resolution on 
$A_{asym}$ estimates 
and found that for an accurate measurement  the resolution
element should be $\leq 0.75$ $h_{50}^{-1}$ kpc. For our 6 galaxies
with redshifts, 0.1 arcsec is 0.8-1.0 $h_{50}^{-1}$ kpc, which may
lead to some underestimation of $A_{asym}$,
 but this is expected to be a small effect 
($<10$ per cent). 

The two relatively undisturbed galaxies, 16V22 and 16V25,
give  $A_{asym}<0.15$, values typical of normal galaxies.
 We find moderately elevated asymmetries 
($A_{asym}\simeq 0.2$) 
for S4 (due to its companion) and all three of the 
ring galaxies, and the highest
asymmetries ($A_{asym}=0.25$--0.5)
 for 16V21, a probable ongoing merger, and the very disturbed S10 and
16V26.
\begin{figure}
\psfig{file=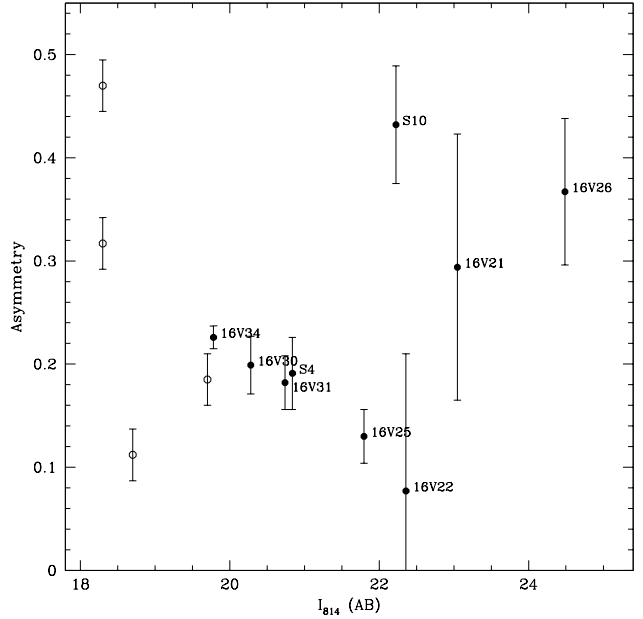,width=90mm}
\caption{Asymmetry parameters of radio-selected galaxies; our sample
of 9 (solid symbols) and the 4 brighter galaxies of Serjeant et al. (2000)
(open symbols).}
\end{figure}
 \begin{table}
\caption{Asymmetry parameters evaluated from WFPC2 $I$-band data for the
9 radio-selected galaxies.}   
\begin{tabular}{lcc}
\hline
Galaxy & $A_{asym}$ & Type\\
S4    & $0.191\pm 0.035$ & E/S0, w companion \\
S10  & $0.432\pm 0.057$ & disk, very asymmetric \\
16V21 & $0.294\pm 0.129$ & disk, merging? \\
16V22 & $ 0.077\pm 0.133$ & $\sim {\rm SBm}$ \\    
16V25 & $0.130\pm 0.026$ & $\sim {\rm SBbc} $  \\
16V26 & $0.367\pm 0.071$ & disk, asymmetric  \\
16V30 & $0.199\pm 0.028$  & bulge w ring \\
16V31 &  $0.182\pm  0.026$ & spiral  w inner/outer rings \\
16V34 & $0.226\pm  0.011$ & disk w ring \\
\hline
\end{tabular}
\end{table}
Figure 4 shows $A_{asym}$ against redshift for our sample and that of 
  Serjeant
et al. (2000).
The mean $A_{asym}$ (0.23) of our sample, the dispersion (0.11) and
the range are all similar to those of the  radio-selected galaxies at
  $z\simeq 0.2$. 

Although even non-interacting irregular
  galaxies (e.g. NGC4449 on Figure 7 of Conselice et al. 2000)
 may have asymmetries of up to  $A_{asym}\simeq 0.26$, these are
  atypical and most galaxies in optically-selected
  samples (e.g. the CFRS/LDSS galaxies on Figure 2 of Sergeant et
  al. 2000) are clustered around lower values of
$A_{asym}\simeq 0.1\pm 0.05$. The results for our
 sample, combined with the $z\sim 0.2$ galaxies, are therefore
  evidence that radio-selected galaxies are on average more
  dynamically disturbed than optically-selected  
galaxies. Futhermore, we see that although
  some of these radio-selected galaxies 
appear normal and have $A_{asym}<0.15$, an even  higher proportion
  (5 of the 13 plotted) have the $A_{asym}>0.25$ asymmetries
  characteristic of ongoing or very recent merging,  or 
  other  major dynamical disruption.

\subsection{Residuals to Fitted Profiles}
Star-forming regions  and asymmetric effects of dynamic disturbance
 can be
highlighted   by subtracting the underlying disk or bulge profile from
 galaxy images, to leave an
image of the residuals. Using {\sevensize IRAF} `mkobjects', we
generate for each galaxy a 2D image of a  symmetrical disk or bulge model
profile, with the same scalelength (from the
fitted profile as shown on Figure 2), and same ellipticity and position
angles (as given by  SExtractor). These model profiles are
subtracted from  the observed galaxies and subtracted, leaving 
the residuals (Figure 5), which are briefly described below.  
\begin{figure}
\psfig{file=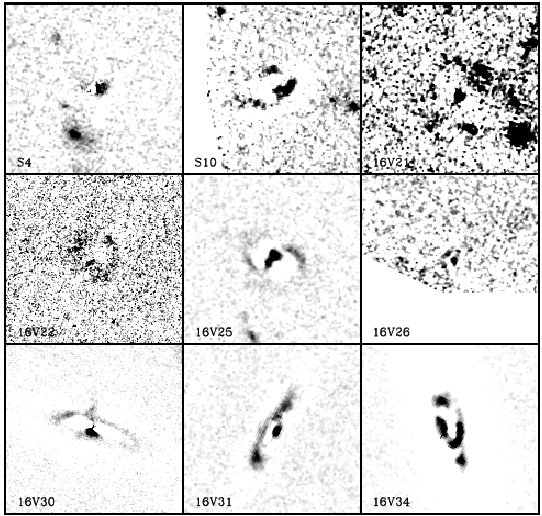,width=90mm}
\vskip-4.0cm
\caption{WFPC2 images of the 9 galaxies with fitted disk and bulge
profiles subtracted, to highlight the residual features.} 
\end{figure}

S4: residual image shows part of the nucleus (incompletely subtracted 
due to its isophotal
twist) and whole of the companion galaxy. S10: bright knots extending W of the nucleus on the N
and S edges of the galaxy. 16V21: bright knots and the possible
companion galaxy and connecting
filament. 16V22: shows a faint ring resulting from the
outer isophotes being rounder than the bar-like inner region. 16V25: bright
residuals consisting of the bar
and arms. The bar is asymmetric, almost
triangular.
 16V26: multiple bright regions. 16V30: shows  part of the slightly off-centre
nucleus, a spur extending to the N, and a large bright ring with a full
circle visible. 16V31: shows
 part of the slightly off-centre nucleus, and a large bright
ring, broken  (obscured by internal extinction?) on one side, 
There is an additional bright region to the N of the nucleus and
within the ring, which on close inspection is seen to lie on a thinner
and fainter inner ring. 16V34: a very bright ring of variable
SB, appearing broken, and an
extended bright region S of the ring.

The most significant new findings from this method are (i) asymmetry
and possible disturbance in the centre of  16V25, and (ii) 
rings in three galaxies. These rings are discussed in
detail below.
\subsection {`Cartwheel'-type Rings}
The `Cartwheel' galaxy (A0035), first described by Fosbury and
Hawarden (1977), is a luminous  $(M_B=-22.45$)
 late-type disk at $z=0.03$, with a large and very prominent  33 $h_{50}^{-1}$
kpc radius ring.
The ring is now known to have been formed
 by the high-velocity passage of a small
companion galaxy through the disk, near the centre, triggering an
expanding density wave with a radial velocity $53\pm 9$ km $\rm s^{-1}$
(Higdon 1996). 
There is also a much
fainter inner ring of radius 7 $h_{50}^{-1}$. Appleton and Marston
(1997) describe 10 other  galaxies of this type with a wide range of
ring radii.

 Collisional ring galaxies are remarkable,
in that the high-velocity impact of  a much less massive galaxy can
 generate a very prolonged
starburst which eventually sweeps through the whole disk, converting most of
the gas to stars.
Despite the Cartwheel's  $\sim 0.6$
$h^{-1}_{50}$ Gyr post-impact age (radius/velocity),
 the SFR remains 
extremely high with almost all the star formation
 concentrated in bright knots on the outer
ring. Higdon (1995, 1996),  on the basis of  Balmer line and
radio/$\rm H\alpha$ ratios, estimate a high extinction of $A_V\simeq 2$
mag for these star-forming regions, and after correcting $H\alpha$ for this
 estimate a SFR of 268
 $h^{-2}_{50}M_{\odot}$ $\rm yr^{-1}$. At this SFR the Cartwheel will 
exhaust its supply of gas
in a further $\sim 0.3$ Gyr.

Our 
WFPC2 sample contains  3 galaxies with rings suggestive of similar
collision histories.
16V31 has the closest resemblance to  the Cartwheel in that a nucleus, inner
and outer rings are all visible, and its radio luminosity and SFR are
very similar. 16V34 differs  in that only one ring is visible, the
ring is
more off-centre and asymmetric (appearing broken), and no nucleus is visible.
 It may simply be that 16V34 is a later-type spiral than 16V31, and
that the colliding dwarf galaxy
impacted further from the centre and 
 at a greater inclination to
the disk plane.
16V30, on the other hand, is an earlier-type galaxy with a very bright
bulge-profile nucleus. The residual image shows a luminous spur
extending N from the nucleus.

Figure 6 shows cross-sections along the major axes of these three
galaxies, with the fitted disk and bulge profiles subtracted (as on
Figure 2). Ring radii are measured from the peaks in 
these cross-sections. On 16V30 (redshift unknown), the ring (peak
$\rm A_1$ to
$\rm A_2$) has radius 1.51
arcsec and is off-centre from the nucleus by 0.17 arcsec. On 16V31 the
outer ring (peak $\rm A_1$ to $\rm A_2$) has radius 1.17 arcsec
 (10.0 $h_{50}^{-1}$
kpc or 1.63 $r_{exp}$) and is off-centre by 0.17 arcsec, and the
 inner ring (peak $\rm B_1$ to $\rm B_2$ has a radius 0.67
arcsec (5.7 $h_{50}^{-1}$ kpc or 0.93 $r_{exp}$)
 and is off-centre by 0.13 arcsec. On
16V34 the ring (peak $\rm A_1$ to $\rm A_2$)
 has radius 0.90 arcsec (7.0  $h_{50}^{-1}$
kpc or 0.71 $r_{exp}$ ) and is more off-centre, by 0.25
arcsec, while the secondary peak B is 0.70 arcsec beyond the A ring.

If the ring expansion velocities are similar to the Cartwheel, the
post-impact ages of 16V31 and 16V34 would be 184 $h_{50}^{-1}$ and 130
$h_{50}^{-1}$ Myr, which are   typical for galaxies with visible 
collisional rings (Bransford et al. 1998).
16V31, with two rings, appears to be  at a similar evolutionary stage
to the Cartwheel, but 16V34 may be at an earlier stage, as there is only
one ring and it is considerably smaller than the visible extent of
the galaxy disk. If, for example, the  starburst does not  run out of
fuel until the ring reaches
$\sim 2 r_{exp}$, 16V34 will be a bright radio source for a further 
 0.2--0.3 Gyr.
\begin{figure}
\psfig{file=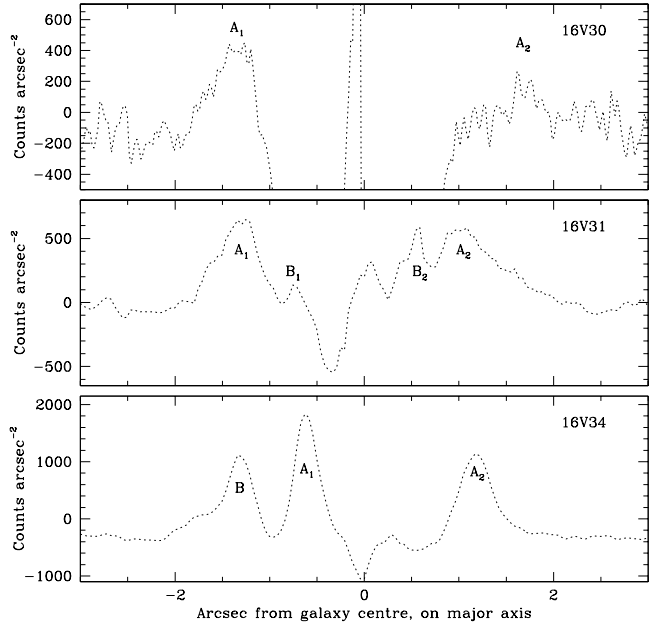,width=90mm}
\caption{Major axis cross-sections of the 3 galaxies with
rings, after subtraction of the fitted disk/bulge profiles. The labels `A'
indicate peaks corresponding to the primary rings. the labels 'B'
indicate the inner ring of 16V31 and the outlying bright region in
16V34.}
\end{figure}. 

\section{The Colours of S4 and S10}
Combining WFPC2 data with the STIS data  for S4 and  S10
 provided $V_{STIS}-I_{814}$ colours, which although 
narrow-based, may still be  
useful in distinguishing young and old stellar populations, especially
at $z>0.5$.  The whole-galaxy colours, as measured in large elliptical
 apertures fitted to $I$-band isophotes
 (including almost all of the flux, and matched in the two passbands)
, are $V_{STIS}-I_{814}=1.39\pm 0.02$ 
for S4 and $0.86\pm 0.04$ for S10. The colour of the
 small companion of S4 was measured in a 0.5 arcsec radius aperture
as  $V-I=0.93\pm 0.05$.
 
Using the `Pegase 2' package of Fioc and Rocca-Volmerange (1997),
 we
 model the  $V_{STIS}-I_{814}$ colours expected for the evolving
 galaxy models considered  in RLK02. The E and S0
 models differ in the initial starburst duration, but both evolve
 passively at $z<2$. The spiral and Irr models have SFRs decreasing
 approximately exponentially and include dust. We also model
 a non-evolving 50 Myr age pure starburst, with and without dust.
Figure 8 shows these models together with the whole-galaxy
colours of S4 and S10, and also
the colour of the companion galaxy to S4,
measured in a 0.5 arcsec radius aperture as $V-I=0.93\pm 0.05$. S4 has
the red colour of a passive galaxy with no indication of ongoing
star-formation, while its disk-type companion and S10 are both
much bluer and consistent with evolving spirals.
\begin{figure}
\psfig{file=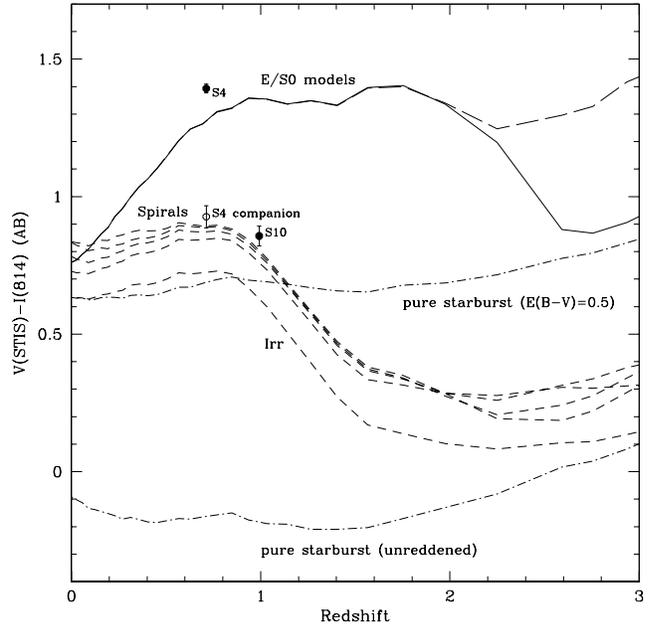,width=90mm}
\caption{$V-I$ colour from WFPC2 and STIS data, against redshift, 
for S10, S4 and its interacting companion; compared with
model colour-redshift relations representing E (solid) S0 (long-dash),
spiral and irregular (short-dash) and starburst (dot-dash) galaxies.}
\end{figure}

Using {\sevensize IRAF} `geomap', the STIS images could be  rebinned
to the pixel grid of the WFPC2 data. Colour profiles were then extracted,
by using `isophote' to measure STIS fluxes on the isophotes already
fitted to the $I$ images, and are shown on Figure 7.
\begin{figure}
\psfig{file=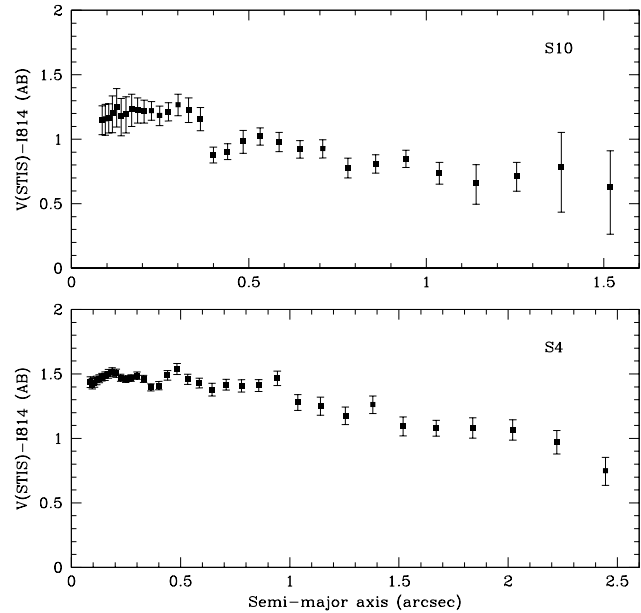,width=90mm}
\caption{$V-I$ colour profile from WFPC2 and STIS images,
 measured on isophotes fitted to the
WFPC2 images, for S4 and S10. Note that the  S4 profile includes the companion
galaxy at $1.5<r<2.5$ arcsec.}
\end{figure}

S10 is significantly non-uniform in colour, with a relatively red
centre, $V-I\simeq 1.2$, which is closer to a passive than a spiral
 model. At $r>0.38$ arcsec there is a sudden transition  
to much bluer colours of  $V-I\simeq 0.7$--0.9,  consistent with an
evolving spiral
 or a moderately reddened ($E(B-V)\simeq 0.5$ mag) starburst. This
 suggests that the galaxy was initially a normal star-forming spiral,
 with a nucleus containing older
 stellar population in its nucleus. This appears to have been 
 greatly disrupted by an
 interaction, triggering extensive starbursting with associated radio
and $\rm [OII]3727$ emission. However, the disk colour remains consistent
with a non-starburst spiral, so the burst may be moderately --
$E(B-V)\leq 0.5$ mag -- dust-reddened.

S4, in contrast, is uniformly red, with passive-evolution colours,
 out to $r\simeq 1$ arcsec. The bluer colour at $r>1.5$ arcsec
is due at least in part to the much bluer companion galaxy. However,
 there is evidence of slight ($\sim 0.2$ mag) 
 bluening at $1<r<1.5$ arcsec, where the radial profile does not indicate any
additional flux from the companion, so there may be a
small  colour
gradient in its outer regions.

\section{Discussion: Nature of the Radioluminous Galaxies}
\subsection{Origin of the Radio Emission}

S4, the most radioluminous, is a giant elliptical,
with a much fainter blue disk companion. 
The colours and SB are consistent with a
passively evolving elliptical and there are no obvious star-forming regions.
We find no indication that it is undergoing the extreme  star formation --
at least $\sim 1000 M_{\odot}$ $\rm yr^{-1}$ -- needed for this to account for 
 its radio luminosity.
The radio emission is probably from an obscured AGN, and it seems plausible
that the interaction with the companion 
triggered the current  nuclear activity. 
 
S10, the second most radioluminous, is a very different object, blue
in colour with 
a very asymmetric disk profile. The bright knots, 
high  optical SB and  strong emission lines are evidence for
 an extensive starburst. 
It has a  much redder nucleus,
suggesting it was initially a spiral. 
Interaction-triggered star-formation
can probably account for the high radio luminosity. 

16V21 is also a highly disturbed disk galaxy with bright
knots, and may be interacting or merging.

16V22 is a sub-$L^*$ galaxy of estimated type $\sim \rm SBm$. 
The SB is near average for a disk
of its redshift, and is not visibly interacting or disturbed.
However, very blue colours,  strong $\rm [OIII]5007 \AA$ emission,
and a
high excitation ratio,  $F({\rm [OIII]})/F({\rm H\beta})=2.93\pm 0.14$ (RLK02)
imply starbursting, or possibly an AGN.
The deep radio sample of Hammer et al. (1995) contained some similar
high-excitation, sub-$L^*$  galaxies, which were of uncertain nature
having line diagnostics near
the HII/Seyfert/LINER divide. In the case of 16V22, we find no sign  of
a central point source on our image, so it seems most likely that
star-formation accounts for the radio emission.

16V25 appears to be a barred $\sim\rm SBbc$ spiral
, with a high disk and central SB.
The radio emission may be due to strong star-formation, possibly
enhanced  by a minor interaction as the central bar appears
asymmetric.
It is also possible that some of the emission is from an AGN, but any
active nucleus must be 
heavily obscured as there is no visible point-source (the redshift was too
high for RLK02 to obtain an excitation ratio).

16V26 appears to be a disturbed disk galaxy, like S10, and on the basis
of its faintness and colours is likely to be at $z>1$.

16V30 could be (i) a very early-type spiral with radio emission from a
collisional
starburst ring, or (ii)  a giant elliptical that has accreted a 
 smaller galaxy, forming a ring. Of these 9 galaxies, it is the second
brightest in $I$ but has the lowest radio flux.
 The redshift is unknown but if 
 the optical luminosity is high ($M_B\sim-22.5$)
the radio emission would be consistent with a normal giant elliptical,
following the Sadler, Jenkins and
Kotanyi (1989) relation.  

16V31 and 16V34 are disks with rings resembling those in the Cartwheel galaxy,
 and on the basis of the high SFR and radio flux 
of the original Cartwheel, it
is highly probable that  collisional starbursting in these rings 
 accounts for their radio luminosities.

\subsection{The Role of Interactions}
Our  HST imaging survey 
confirms the high incidence of interactions amongst
faint radio-selected galaxies, as previously 
reported by e.g.  Windhorst et
al.  (1995) and Serjeant. et al.  (2000), with the strong radio
emission resulting from starbursting in the majority of galaxies (at
least 6 of this sample of 9) and obscured AGN in a significant minority
(at least 1/9 here). 
 Probably the  most interesting new findings are
 the Cartwheel-type rings in at least 2 and possibly 3
of these galaxies. 

These rings, together with  the asymmetry of  S10  and 16V26, imply
that a rather large fraction, 5/9, of the sample are post-encounter interacting
galaxies, whereas  only two appear to be in pre-encounter pairs. 
 This can be compared with the ULIRGs ($L_{60\rm \mu
m}>10^{12}h^{-2}_{50}L_{\odot}$ galaxies) of  Clements et
al. (1996),  of which 55/60 are visibly disturbed or merging, and
double nuclei could be resolved in 28, i.e. about equal numbers are
observed    before    and   after    nuclear
coalescence/collision.
 A high post-encounter fraction amongst 
radio-selected galaxies might be expected on the basis of
a model of  Lisenfeld et al. (1996), in which the synchrotron
radio emission from
a starburst remains strong for at least $\sim 80$ Myr after 
star formation ceases. This also means that  galaxies that  have already
been starbursting for $>80$ Myr would have higher radio
luminosities per unit SFR, and higher radio/FIR ratios, than 
near the
onset of star formation.

16V31 has strong $\rm H\delta$ absorption which could be fit by 
 a Delgado et al. (1999) post-starburst model at age $\sim 27$ Myr, but 
 could also result from `age-dependent extinction', with
starbursting ongoing for 
$>10^8$ yr with only the most recent
star-formation ($<10^7$ yr)   obscured by dust (Poggianti, Bressan and
Franceschini 2001). The LRIS spectroscopy of RLK02 did not
provide the line ratios, e.g. the Balmer decrement, needed to
 distinguish these possibilities. However,
the  close resemblance to the Cartwheel is an important
clue, as the Cartwheel is known  to have a very high current SFR combined with
$A_V\simeq 2$ mag extinction of the starburst emission lines. A similar
picture is seen for the other 8
 collisional ring galaxies studied by Bransford et al. (1998),
including two with strong Balmer absorption lines. 

 This is significant, as prolonged starbursting with age-dependent
 extinction can (in addition to producing strong Balmer lines),
 explain
 the very low emission line to radio flux ratios of 
$\rm \mu Jy$ radio-selected galaxies in general (see RLK02). 

 Collisional ring
galaxies may prove to be quite prominent in deep  $\rm \mu Jy$
radio surveys,  as they are more numerous at higher redshifts 
(Lavery et al. 1996) 
and the long synchrotron lifetime  would favour the radio-detection
of post-encounter, $\geq 10^8$ yr age,  prolonged  starbursts,
 of which Cartwheel-type galaxies are  prime
 examples. 

\subsection*{Acknowledgements}Based on observations with the NASA/ESA
Hubble Space Telescope obtained at the Space Telescope Science Institute, which is operated by the Association of Universities for Research in Astronomy, Incorporated, under NASA contract NAS5-26555.
. DCK was supported by an NSF PYI
grant AST-8858203 and a research grant from UC Santa Cruz. NR is currently
supported by a PPARC research associateship. 
We thank Matthew Bershady and Christopher Conselice for providing their
software for calculating galaxy asymmetries. 
 
\end{document}